\documentclass[12pt]{article}
\usepackage{latexsym}
\newcommand{\be}{\begin{equation}}
\newcommand{\ee}{\end{equation}}
\newcommand{\bq}{\begin{eqnarray}}
\newcommand{\eq}{\end{eqnarray}}
\newcommand{\ids}{\int d^{2}\!\sigma}
\begin{document}
\begin{titlepage}
\today          \hfill 
\begin{center}
\hfill    LBNL-44122 \\
 \hfill    UCB-PTH-99/35 \\
\hfill  hep-th/9908176
\vskip .5in

{\large \bf Conformal Models of Thirring Type\\
and the Affine Virasoro Construction}
\footnote{This work was supported in part by the Director, Office of 
Science, Office of High Energy and Nuclear Physics, Division of 
High Energy Physics of the U.S. Department of Energy under Contract 
DE-AC03-76SF00098 and in part by the National Science Foundation under 
grant PHY-95-14797.}

\vskip .5in
Korkut Bardakci\footnote{e-mail:kbardakci@lbl.gov}

{\em Department of Physics\\
University of California at Berkeley\\
   and\\
 Theoretical Physics Group\\
    Lawrence Berkeley National Laboratory\\
      University of California\\
    Berkeley, California 94720}
\end{center}

\vskip .5in

\begin{abstract}
We investigate a class of models in 1+1 dimensions with four fermion
interaction term. At each order of the perturbation expansion, the models
are ultraviolet finite and Lorentz non-invariant. We show that for certain
privileged values of the coupling constants, Lorentz symmetry is restored,
and indeed the model turns out to be conformally invariant. This
phenomenon is both quantum mechanical and non-perturbative.
\end{abstract}
\end{titlepage}

\newpage
\renewcommand{\thepage}{\arabic{page}}
\setcounter{page}{1}
\noindent{\bf 1. Introduction}
\vskip 9pt

Conformal models have played an important role both in the statistical 
mechanics of lower dimensional systems and in the construction of viable
string theories. In the absence of a systematic approach so far
 for the classification and the construction of conformal theories,
various special models have been proposed and applied to string
 theory [1,2].
Among these are the conformal models based on the affine Virasoro
 construction
[3,4], which are generalizations of the original Sugawara model in
terms of currents that satisfy an affine algebra. Apart from some isolated
cases [5], this construction has so far not found widespread application
in string theory. Among the reasons for this is the lack of a good
understanding of the local field theory that is at the basis
 of this construction.
Several formulations of the model based on various generalized sigma model
type actions have been proposed [6,7,8], but more work is needed to make
progress along this direction.

A different approach to the same problem is to start with various 
generalizations of the Thirring model and investigate possible
non-trivial fixed points in the coupling constant space. This approach
is motivated by the observation that the interaction in these models is
of the form current$\times$current, which is very suggestive of the affine
Virasoro construction. The original Thirring model has already been used
in string compactification [9]. In their classical work, Dashen and
Frishman [10] showed that a non-abelian
 generalization of the Thirring model symmetric
under a Lie group is conformally invariant for certain quantized values of
the coupling constant, and at these conformal points, the stress tensor 
admits an affine- Sugawara
 construction. More recent work [11,12,13] suggests
the possible existence of more general non-symmetric fixed points 
in the coupling constant space.

In this paper, we shall investigate a different generalization of the
Thirring model. The models in question have some unusual properties.
Since the coupling constants are dimensionless, one would expect the
appearence of the usual renormalizable divergences in perturbation theory.
Instead, it turns out that  each order of perturbation expansion is
ultraviolet finite.
 Another surprising feature is connected with Lorentz invariance.
 Superficially, the interaction term in these models seems to
 violate Lorentz
invariance on the world sheet,  and if true, this would disqualify them
from being of  use in string theory.\footnote{It was also suggested
in reference [4] that affine Virasoro constructions are surrounded
by non-Lorentz invariant theories.}
 We will however show that, if the
coupling constants in the interaction term satisfy the
 Virasoro master equation or simply the master equation
 [4], the corresponding
 model is conformally invariant,
and the stress tensor is given by the affine Virasoro construction.
 It then
follows trivially that, since the conformal
group includes Lorentz transformations,
 contrary to the appearances,
 the model is also Lorentz invariant. We would like to emphasize that
Lorentz invariance is realized non-perturbatively only at the points in
 the
coupling constant space that satisfy the master equation; individual terms
in the perturbation expansion violate this symmetry.
 It is also of some interest to
determine the transformation properties of the various fields
 that appear in
the model under the Lorentz group, since in general fields could transform
 non-lineary in a complicated fashion.
We have investigated the transformation properties of the basic fermion
fields and the currents, which are bilinear composites of the fermions.
It turns out that the fermions transform linearly, with, however an
anomalous coefficient for the ``spin'' term.
The currents also transform linearly, with a different coefficient for
the spin term. Again, these simple transformation properties hold
only at the conformal points in the space of coupling constants.

A question of some interest is whether it is possible to find a
model based on bosonic fields with the properties discussed above;
namely, lack of classical Lorentz symmetry and its quantum
mechanical restoration. One can in fact construct such a model
directly by bosonizing the fermionic field theory we are going to
introduce in the next section. The methods of bosonization are well
known, and they have already been applied to similar Thirring type
models [11,12]. It is then quite easy to adopt them to the
problem at hand. However, the resulting theory has a complicated
structure, and it seems quite difficult to analyze quantum
mechanical Lorentz invariance in this formulation. This is the
main reason why we have based our exposition on the fermionic
version of the theory. 

The paper is organized as follows. In section 2, we will introduce the
model and argue for the lack of ultraviolet divergences in perturbation
theory. In section 3, we will review free fields and the affine
Virasoro construction. In section 4, we will show that if the coupling
 constants satisfy the master equation, the model is conformally
 invariant.
 The demonstration is 
based on the calculation of the operator form of the stress tensor in
the standard interaction representation. In section 4, the Lorentz
transformation properties of the fermions and the currents will be
determined. Finally, the last section will summarize our conclusions.
\vskip 9pt
\noindent {\bf 2. The Model}
\vskip 9pt

The model is based on the following action:
\be
I=I_{0}+I',
\ee
where,
\be
I_{0}=\frac{i}{2}\ids\,\bar{\psi}^{a}\gamma^{\mu}\partial_{\mu}\psi^{a},
\ee
and,
\be
I'= \ids\,\left(g\, c_{ij}\, J_{+}^{i} J_{+}^{j}
+g'\, c'_{ij}\, J_{-}^{i} J_{-}^{j}\right).
\ee
In these equations, $\psi^{a}$'s are two component Majorana spinors in
1+1 dimensions. As is usual in string theory, the time coordinate is
denoted by $\tau$ and the space coordinate by $\sigma$. Again in keeping
with the string usage, we will take $\sigma$ to be compact and to
 range from
0 to $2\pi$, although this is not important for most of the subsequent
development. The coupling constants $c_{ij}$ and $c'_{ij}$
 that appear in the interaction are real and symmetric in i and j.
 We have also introduced two
redundant  constant g and g' for later convenience. The currents
 $J^{i}_{\pm}$
of definite chirality are constructed from chiral fermions 
 $\psi^{a}_{\pm}$:
\be
J^{i}_{\pm}=\frac{1}{2}\psi^{a}_{\pm}\,
\lambda^{i}_{ab}\,\psi^{b}_{\pm},
\ee
where $\lambda^{i}$ are matrices which act on the internal space labeled
by a and b. They satisfy the commutation relations
$$
[\lambda^{i},\lambda^{j}]=if^{ijk}\lambda^{k},
$$
and generate some Lie algebra. In what follows, we will take this algebra
to be semi-simple with the metric given by identity, so that there will
be no need to distinguish between upper and lower indices.

We shall now argue that the perturbation expansion for these models is
ultraviolet finite. To start with,
 it is  clear that
the $ +$ and $-$ chiralities never mix, and therefore they can be
 considered
seperately in the perturbation expansion. As a simple example,
 consider the one loop contribution to the $+$ chirality fermion-fermion
 scattering. If p is the total external momentum, suppressing all the
dependence on the internal space indices and overall constants,
one encounters a potentially divergent integral of the form
\bq
M&\cong&\int\frac{d^{2}k}{(2\pi)^{2}}\frac{(k_{0}-k_{1})(p_{0}-k_{0}
-p_{1}+k_{1})}{k^{2} (p-k)^{2}}\nonumber\\
&=& i\int_{0}^{1} d\alpha \int_{0}^{\infty} k dk \int_{0}^{2\pi} d\theta
\frac{\left( i k e^{i\theta}+ \alpha (p_{0}-p_{1})\right)
\left((1- \alpha)(p_{0}-p_{1}) -i k e^{i\theta}\right)}
{\left(k^{2}+ (\alpha^{2}- \alpha) p^{2}\right)^{2}}\nonumber\\
&=& - \frac{i}{2} \frac{(p_{0}-p_{1})^{2}}{p^{2}}.
\eq
The integral, which superficially appeared to be logarithmically
divergent, is actually convergent. This is because  the k dependent
terms in the numerator on the second line of the equation, which would
normally lead to a divergent integral,  all vanish after the
integration over the angle $\theta$. Since the fermions in the
Feynman graph all have positive chirality, the propagators always carry
 a factor $k_{0}-k_{1}$, which after Euclidean rotation turns into 
 $i k\, e^{i\theta}$. The important point is that factors of k in the 
numerator  always appear in the combination $k\,e^{i\theta}$
 and,  as a result, all of the $\theta$
dependent  terms vanish upon integration. But since these are the only
possible divergent terms, the integral must be finite.
 It is easy to see that this argument
works also for higher order graphs, and it follows that all of them are
finite. A similar argument, with a change of the sign of $\theta$,
shows that all of the graphs with negative chirality fermions are also
finite. The only potentially divergent graphs are the ones that contain
both positive and negative chirality fermions, but because of the form
of the interaction (eq.(3)), there are no graphs of this type.

The price paid for the finiteness of the model is the loss of Lorentz
invariance, at least in the perturbation expansion. 
  Lorentz invariant interactions must conserve chirality,
 which is not the case in our model. One can check the breakdown of 
Lorentz invariance explicitly
in the case fermion-fermion scattering process discussed above.
 Higher order
graphs for this process will yield an answer proportional to a factor
of the form 
$$
\left(\frac{(p_{0}-p_{1})^{2}}{p^{2}}\right)^{n}
$$
where the integer n will depend on the order of perturbation theory.
Since the above factor scales under Lorentz transformations, different
orders in perturbation expansion will have different Lorentz
 transformation
properties. This only means, however, that there is no Lorentz invariance
for arbitrary values of the coupling constants. In section 4, we will show
that, for special values of the coupling constants, Lorentz invariance is
restored.
\vskip 9pt
\noindent{\bf 3. Free Field Constructions}
\vskip 9pt

In this section, we will review the free field limit of the model,
with g set equal to zero, and introduce the
affine Virasoro construction as a preparation for the next section.
 The free fermions of definite chirality, $\psi^{a}_{0,\pm}$, depend
on the coordinates through the combinations $\sigma\mp \tau$,
and they satisfy the following  commutation  relations:
\be
[\psi^{a}_{0,\pm}(z),\psi^{b}_{0,\pm}(z')]=
\delta_{a,b}\, \delta(z-z').
\ee
In this equation, z stand for $\sigma-\tau$ for the $+$ components and
for $\tau+\sigma$ for the $-$ components.
Free currents $J^{i}_{0,\pm}$ are constructed from free fermions as in
eq.(4), with a normal ordering prescription. They satisfy the commutation
relations
\be
[J^{i}_{0,\pm}(z),J^{j}_{0,\pm}(z')]= i f_{ijk}\, \delta(z-z')
J^{k}_{0,\pm}(z)
-\frac{i\kappa}{2\pi}\,\delta_{i,j}\, \delta'(z-z'),
\ee
where z has the same meaning as before.
 The constant $\kappa$ is the coefficient of the central term, and,
 given the representation matrices $\lambda^{i}$, it
 can easily be calculated.

Another quantity of interest is the stress tensor of the free field
theory. Since there is no mass term, the stress tensor is traceless, and
and the two independent components can be conveniently taken to be 
$$
T_{0,\pm}=\frac{1}{2}\left(T^{0,0}_{0}\pm T^{0,1}_{0}\right).
$$
In terms of free fermions, they are given by the normal ordered
expression
\be
T_{0,\pm}=\pm\frac{i}{4}\left(:\partial_{\sigma}\psi^{a}_{0,\pm}\,
\psi^{a}_{0,\pm}:-:\psi^{a}_{0,\pm}\,\partial_{\sigma}\psi^{a}_{0,\pm}:
\right).
\ee
$T_{0,\pm}$ satisfy the conservation equations
\be
(\partial_{\tau}+\partial_{\sigma})T_{0,+}=
(\partial_{\tau}-\partial_{\sigma})T_{0,-}=0,
\ee
and therefore they are functions of only the variables $\tau \mp \sigma$
respectively. We also note that their commutators generate the Virasoro
algebra:
\be
[T_{0,\pm}(z),T_{0,\pm}(z')]= \pm i \delta'(z-z')\left(T_{0,\pm}(z) +
T_{0,\pm}(z')\right) +\frac{i c_{0}}{2\pi}(\partial^{3}_{z} +\partial_{z})
\delta (z-z'),
\ee
where $z= \sigma\mp \tau$ as before. The numerical value of the
coefficient $c_{0}$ of the central term will not be needed.

After this discussion of free fields, we will briefly review the affine
Virasoro construction and the master equation. The affine Virasoro 
construction is an ansatz for the stress tensor in terms of free currents:
\bq
L_{+}(z)& =& c_{ij}\,:J^{i}_{0,+}(z) J^{j}_{0,+}(z):,\nonumber\\
L_{-}(z)&=& c'_{ij}\,:J^{i}_{0,-}(z) J^{j}_{0,-}(z):,
\eq
where the double dots imply normal ordering in order to have a well
 defined product of the currents.
The basic idea is to require $L_{\pm}(z)$ to satisfy the Virasoro algebra
\be
[L_{\pm}(z),L_{\pm}(z')]= \mp i \delta'(z - z') \left( L_{\pm}(z)
 + L_{\pm}(z')\right)
+\frac{i c}{2\pi}(\partial^{3}_{z}+\partial_{z})\delta(z-z'),
\ee
given that the currents satisfy the commutation relations of eq.(7).
It can be shown that [4] this leads to the following equation
(the master equation) for the
constants $c_{ij}$:
\be
2\pi\, c_{ij}=
 2\kappa c_{ik} c_{kj} - f_{kli} f_{k'l'j} c_{kk'} c_{ll'}
 - f_{kk'l} f_{ll'j} c_{ik} c_{k'l'} - f_{l'k'l} f_{kli} c_{kl'}
c_{k'j},
\ee
with a similar equation for $c'_{ij}$.
A large number of solutions to  eq.(13) with real and symmetric
 c's are known [4]. Any one of them would be satisfactory for
 our purposes.

Another commutation relation that will be needed in the future is
\be
[T_{0,\pm}(z), L_{\pm}(z')]= \pm i \delta(z-z') L'_{\pm}(z)
\mp 2 i \delta'(z-z') L_{\pm}(z'),
\ee
which follows from free field commutation relation after the use of
eqs.(8) and (11).

The use of the same symbols $c_{ij}$ and $c'_{ij}$
 in both the above equation and in
 eq.(3) was not accidental; from now on we will fix the c's
as well as the $c'$'s that appear in the
interaction term in eq.(3) to be a real and symmetric
 solution to  their respective master equations. For the time
being, g and g' are arbitrary; later, they will also be fixed.

\vskip 9pt
\noindent{\bf 4. The Interaction}
\vskip 9pt

After having fixed the constants in the interaction term, we are going to
study the model in the interaction representation. Our goal is
 to establish
the conformal invariance of the model. Since the model is translation
invariant, one can easily construct
the translation operators $P^{0}$ and $P^{1}$
 by the usual Noether procedure.
 However, since
the model is not manifestly Lorentz invariant, this stress tensor is not
symmetrical, and the existence of the generators of the Lorentz group, let
 alone the conformal group, is problematic. Since we can no longer employ
to the Noether construction, we will instead show that the stress tensor
can be determined uniquely by appealing to the following principles:\\
a) The stress tensor should be local function of the coordinates.\\
b) It should be symmetric and traceless.\\
c) The components
$$
T_{\pm}= \frac{1}{2}(T^{0,0}\pm T^{0,1})
$$
 should satisfy the conservation equations (9).\\
d) The energy and momentum operators  should be given by the
standard expressions
\be
P^{\pm}=\frac{1}{2}(P^{0} \pm P^{1})=\int d\sigma T^{\pm}.
\ee
 The conditions we have listed above refer to operators in
the Heisenberg picture. However, for technical reasons, we have
found it convenient to go through the intermediate step of the interaction
representation. The great advantage of this picture is that all the
manipulations involve only free fields.

We remind the reader of a few well known facts about the interaction
representation. In this
picture, the field $\psi$ is taken to be the free field $\psi_{0}$, and
the states satisfy the Schroedinger equation
\be
i \partial_{\tau}|\tau\rangle = H_{I}(\tau)|\tau\rangle,
\ee
where the interaction Hamiltonian $H_{I}$ is given in terms
 free currents by
\bq
H_{I}(\tau)&=& - \int d \sigma\Big(g\, c_{ij}\,:J^{i}_{0,+}(\tau -\sigma)
J^{j}_{0,+}(\tau -\sigma):\nonumber\\
&+& g'\, c'_{ij}\,:J_{0,-}^{i}(\tau +\sigma)
J^{j}_{0,-}(\tau +\sigma):\Big)\nonumber\\
&=& -\int d \sigma\left(g\,  L_{+}(\tau -\sigma) +g'\, L_{-}(\tau +\sigma)
\right)\nonumber\\
&=& -\int d\sigma \left( g\, L_{+}(\sigma) + g'\, L_{-}(\sigma)\right).
\eq
Actually, we are interested in the fields expressed in the Heisenbeg
picture, but we find it advantageous to rewrite them in terms free
fields of the interaction picture. For this purpose, we need the Dyson
operator $U(\tau,0)$, which governs the time development of the states
in the interaction representation:
$$
|\tau\rangle = U(\tau,0)|\tau=0\rangle.
$$
From its definition, this operator satisfies
\bq
i\partial_{\tau}U(\tau,0)&=& H_{I}(\tau) U(\tau,0),\nonumber\\
U(\tau=0,0)&=&1,
\eq
where the Heisenberg and the interaction pictures are taken to coincide
at time $\tau=0$. In our case, the above equation is easily integrated
since $H_{I}(\tau)$ given by eq.(17) is $\tau$ independent:
\be
U(\tau,0)= \exp(- i \tau H_{I}).
\ee

A general field operator $\phi(\tau,\sigma)$ in the Heisenberg picture
can be expressed in terms of the same operator $\phi_{I}(\tau,\sigma)$,
 in the interaction picture by the equation
\be
\phi(\tau,\sigma)= U^{-1}(\tau,0) \phi_{I}(\tau,\sigma) U(\tau,0).
\ee
$\phi_{I}$ is either a free field, or for a composite operator
like the currents or the stress tensor, it is a product of free fields.
In what follows, we will specify various operators of interest in the
interaction picture in terms of free fields, and attach to them an index
$I$ in order to distinguish them from the Heisenberg operators, which will
be free of this index. The latter can then be constructed explicitly
through eqs.(19) and (20).

Having gotten these preliminaries out of way, we are ready to specify
the combination of components $T^{\pm}_{I}$ in the interaction picture.
The unique solution that is local and that satisfies the energy and
momentum conditions of eq.(15) is,
\bq
T_{I,+}&=& T_{0,+}(\sigma -\tau) -\frac{g}{2} L_{+}(\sigma -\tau)
-\frac{g'}{2} L_{-}(\sigma +\tau),\nonumber\\
T_{I,-}&=& T_{0,-}(\sigma +\tau)-\frac{g}{2}L_{+}(\sigma -\tau)
-\frac{g'}{2} L_{-}(\sigma +\tau).
\eq
As usual, the $+$ components of free fields depend only on the variable
$\sigma -\tau$ and the $-$ components on the variable $\sigma +\tau$.
The remaining components of $T_{I}$ can then be solved for using the 
symmetry and the zero trace condition, and therefore, these conditions are
automatically satisfied. 

Now that we have the stress tensor in the interaction picture,
  we can translate it into the Heisenberg picture. It is easy to
carry out the calculation explicitly. For example,
\bq
i\partial_{\tau}\left(U^{-1}(\tau,0) L_{+}(z) U(\tau,0)\right)
&=&U^{-1}(\tau,0)[L_{+}(z), H_{I}] U(\tau,0)\nonumber\\
&=& i g L'_{+}(z),
\eq
where, in the last step, the expression for $H_{I}$ in terms of $L_{\pm}$
(eq.(17)) and
 the commutation relations (12) were used.
This equation, and a similar one for $L_{-}(z)$, have the solutions
\bq
U^{-1}(\tau,0) L_{+}(z) U(\tau,0)&=& L_{+}(z+ g\tau),\nonumber\\
U^{-1}(\tau,0) L_{-}(z) U(\tau,0)&=& L_{-}(z- g'\tau).
\eq
Proceeding in the same fashion, we have,
\be
i\partial_{\tau}\left(U^{-1}(\tau,0) T_{0,+}(z) U(\tau,0)\right)
= ig\,U^{-1}(\tau,0) L'_{+}(z) U(\tau,0) =ig\,L'_{+}(z+g\tau).
\ee
This equation, and the corresponding one for $T_{0,-}$, have the solutions
\bq
U^{-1}(\tau,0) T_{0,+}(z) U(\tau,0)&=& L_{+}(z+g\tau) - L_{+}(z)
+ T_{0,+}(z),\nonumber\\
U^{-1}(\tau,0) T_{0,-}(z) U(\tau,0)&=& L_{-}(z-g'\tau) - L_{-}(z)
+ T_{0,-}(z).
\eq
Putting everything together, we can convert the stress tensor in the
interaction picture given by eq.(21) into the Heisenberg picture:
\bq
T_{+}(\tau,\sigma)&=& (1 -\frac{g}{2}) L_{+}(\sigma-\tau+g\tau)
- L_{+}(\sigma -\tau) -\frac{g'}{2} L_{-}(\sigma+\tau -g'\tau)\nonumber\\
&+& T_{0,+}(\sigma -\tau),\nonumber\\
T_{-}(\tau,\sigma)&=& (1- \frac{g'}{2}) L_{-}(\sigma+\tau -g'\tau)
+ L_{-}(\sigma+\tau) -\frac{g}{2} L_{+}(\sigma -\tau +g\tau)\nonumber\\
&+& T_{0,-}(\sigma +\tau).
\eq

According to the conservation equations, $T_{+}$ should be a function of
only $\sigma -\tau$ and $T_{-}$ should be a function of only
 $\sigma +\tau$.
This requirement fixes the coupling constants to be the following four
combinations:
\be
g= 0, 2 \;\;\; g'=0, 2.
\ee
The zero values for the coupling constants correspond to the
 trivial free field
solutions. We exhibit below the solution $g=g'=2$:
\bq
T_{+}(\tau,\sigma)&=& T_{0,+}(\sigma -\tau) - L_{+}(\sigma -\tau)
- L_{-}(\sigma -\tau),\nonumber\\
T_{-}(\tau,\sigma)&=& T_{0,-}(\sigma +\tau) -L_{-}(\sigma +\tau)
- L_{+}(\sigma +\tau).
\eq

As a further confirmation of conformal invariance, one can easily show
that $T_{\pm}$ satisfy the Virasoro algebra (eq.(12)). It is also
easy to check that the model has Poincare invariance. The single Lorentz
generator is given by the standard expression
\be
M(\tau)= \int d \sigma \left((\sigma -\tau)T_{+}+(\sigma+\tau)
 T_{-}\right),
\ee
and, using the Virasoro algebra, the Poincare commutation relations
$$
[M,P^{\pm}]= \pm i P^{\pm},
$$
are easily verified. Here we have the interesting situation of the
restauration of Lorentz symmetry as a result of quantum effects
 in a model that violates this symmetry classically. This is a
non-perturbative phenomenon that happens only for certain fixed values
of the coupling constants.

We have just seen that the stress tensor has a simple exact expression
in terms of free fields, even though the model is interacting. This
simplification only occurs at the conformal points, with c's fixed
by the master equation (13) and g's fixed by (27). Another set of fields
that are
exactly calculable in terms of free fields are $L_{\pm}$ (see eq.(23)).
However, as far as we know, no other fields enjoy this property
 even when the model is conformal.

\vskip 9pt

\noindent{\bf 5. Lorentz Transformations}

\vskip 9pt

It is of some interest to find the transformation properties of
 the fermion
fields and the currents under Lorentz transformations. In what follows,
we will set $g=g'=2$ and
 focus on the $+$ chirality fields; the calculation for the $-$
 chirality fields is entirely analogous. To find the transformation law of
 $\psi_{+}$, for example, one has to compute its commutator
  with the Lorentz generator $M$ of eq.(29). It is
easiest to do this calculation at $\tau=0$; at this point, $U=1$, and the
Heisenberg and the interaction pictures coincide. One can then carry out
the computation in the interaction picture using free fields. The
equal time commutator of the fermion field with the stress tensor
 has the form
\bq
&&[\psi_{0,+}(\tau,\sigma), T_{I,+}(\tau,\sigma')]=
 \delta(\sigma -\sigma')
A(\tau,\sigma)\nonumber\\
&& + \delta'(\sigma -\sigma')\left( B(\tau,\sigma) +
B(\tau, \sigma')\right),
\eq
where $A$ and $B$ will be calculated below. Given this result, the
commutators of $\psi_{+}$ with the Poincare generators at $\tau=0$
 are easily found to be
\bq
[\psi_{0,+}(0,\sigma),P^{+}]&=& [\psi_{0,+}(0,\sigma), \int d\sigma'
T_{I,+}(0,\sigma')]\nonumber\\
&=& A(0,\sigma) + \partial_{\sigma}B(0,\sigma),\nonumber\\
{}[\psi_{0,+}(0,\sigma), M(0)]&=& [\psi_{0,+}(0,\sigma), \int d \sigma'
\sigma'\, T_{I,+}(0,\sigma')]\nonumber\\
& =& \sigma\, A(0,\sigma) + B(0,\sigma)
+\sigma\, \partial_{\sigma}B(0,\sigma),
\eq
leading to the result
\be
[\psi_{0,+}(0,\sigma),M(0)]= \sigma\, [\psi_{0,+}(0,\sigma), P^{+}]
+B(0,\sigma).
\ee
The first term on the right comes from the transformation of the
coordinates, since
$$
\frac{i}{2}(\partial_{\tau}\mp \partial_{\sigma}) \psi= [\psi, P^{\pm}].
$$
The second term, $B$, is then the spin transformation term. This is also
the term that determines the conformal weight of $\psi$.

It remains to find what $A$ and $B$ are. From eq.(21), we see that we need
 the commutators of $\psi_{+}$ with $T_{0,+}$ and $L_{+}$.
The first commutator contributes the following terms to $A$ and $B$:
\be
A_{1}= -\frac{3 i}{4} \partial_{\sigma}\psi_{0,+},\;\;
B_{1}= -\frac{i}{4} \psi_{0,+}.
\ee
Next, we need the commutator of $\psi_{+}$ with $L_{+}$.
  This part of the computation is a bit more involved, since
 the expression
for $L_{+}$ given by (11) has to be regularized, and for this purpose,
we found it convenient to use the operator product expansion.
 We will first
calculate the OPE of $\psi_{0,+}(z)$ with $L_{+}(z')$,
 where $z=\sigma -\tau$,
 and then convert
the result into the equivalent result for the commutator.
 In this
approach, it is natural to regulate $L_{+}$ by point splitting. We let
\be
L_{+}(z)\rightarrow c_{ij}\,J_{0,+}^{i}(z+\epsilon) J^{j}_{0,+}(
z-\epsilon),
\ee
subtract the term singular in $\epsilon$, and let $\epsilon\rightarrow 0$
at the end. In this case, because of the symmetry of $c_{ij}$ in i and j,
the singular term does not contribute, so one can forget about it.

Starting with the basic OPE
\be
\psi_{0,+}(z)\,J^{i}_{0,+}(z')\cong \frac{1}{4\pi i}\, \frac{1}{z - z'}
\left(\lambda^{i}\psi_{0,+}(z) + \lambda^{i} \psi_{0,+}(z')\right),
\ee
from(34), we have,
\bq
&&\psi_{0,+}(z)\,L_{+}(z')\cong \frac{c_{ij}}{4\pi i}\,
\frac{1}{z -z' -\epsilon}\,
\lambda^{i}\left(\psi_{0,+}(z) + \psi_{0,+}(z'+\epsilon)\right)
J^{j}_{0,+}(z' -\epsilon)\nonumber\\
&& + \frac{c_{ij}}{4\pi i}\,\frac{1}{z -z'+\epsilon}\,
 J^{i}_{0,+}(z'+\epsilon)
\lambda^{j}\left(\psi_{0,+}(z) +\psi_{0,+}(z' -\epsilon)\right).
\eq
One has to apply the OPE (35) once more to the products of the form
$J_{0,+}\psi_{0,+}$ on the right hand side of this equation. After that,
the limit $\epsilon\rightarrow 0$ can  be taken without encountering
any singularities:
\bq
&&\psi_{0,+}(z)\,L_{+}(z')= 
 \frac{c_{ij}}{8\pi^{2}}\frac{1}{z - z'}
\lambda^{i}\lambda^{j}\left(\psi'_{0,+}(z) +\psi'_{0,+}(z')\right)
\nonumber\\
&&- \frac{c_{ij}}{8\pi^{2}}\,\frac{1}{(z - z')^{2}}\lambda^{i}
\lambda^{j}\left(\psi_{0,+}(z) +\psi_{0,+}(z')\right)\nonumber\\
&&+ \frac{c_{ij}}{2\pi i}\, \frac{1}{ z- z'}\left(:J^{i}_{0,+}(z)
\lambda^{j}
\psi_{0,+}(z): + :J^{i}_{0,+}(z')\lambda^{j}\psi_{0,+}(z')\right).
\eq
The double dots around the composite operators indicate that these terms
have been regulated by subtracting the short distance singularity. By 
applying the usual translation
$$
\frac{1}{2\pi i}\,\frac{1}{z - z'}\rightarrow \delta(z - z'),
$$
The contributions from the commutator of $\psi_{0,+}$ with $L_{+}$ to
$A$ and $B$ are
\bq
A_{2}&=& c_{ij}\left(\frac{i}{2 \pi}
\lambda^{i}\lambda^{j}\partial_{\sigma}\psi_{0,+}+4 :J^{i}_{0,+}
\lambda^{j}\psi_{0,+}:\right),\nonumber\\
B_{2}&=&\frac{i}{2\pi}c_{ij}\,\lambda^{i}\lambda^{j}\psi_{0,+}.
\eq
The total values of $A$ and $B$ are obtained by adding up these results:
$$
A= A_{1}+ A_{2},\;\;\;B= B_{1}+ B_{2},
$$
and we finally have the following transformation law for $\psi_{+}$:
\bq
&&{}[\psi_{+}(\tau,\sigma),M(\tau)]=
\frac{i}{2}(\sigma -\tau)(\partial_{\tau} -\partial_{\sigma})
\psi_{+}\nonumber\\
&& -\frac{i}{4}\left(1  -\frac{2}{\pi}c_{ij}\,\lambda^{i}
\lambda^{j}\right)\psi_{+}.
\eq

A similar calculation for the current $J_{+}^{k}$ gives
\bq
&&[J_{+}^{k}(\tau,\sigma), M(\tau)]=\frac{i}{2}(\sigma -\tau)
(\partial_{\tau} -\partial_{\sigma}) J^{k}_{+}\nonumber\\
&&-\frac{i}{2} J^{k}_{+}+
\frac{i}{\pi}\left(\kappa\,c_{km} +\frac{1}{2} c_{ij}\,f_{ikl}
\,f_{jml}\right) J^{m}_{+}.
\eq
These equations show that both the fundamental fermion and the current
transform linearly, but they
have anamolous spin terms. This is equivalent to the
 existence of anomalous
conformal dimensions. Fields with definite transformation properties
are obtained by diagonalizing the matrices that appear in these equations.
\vskip 9pt

\noindent{\bf 6. Conclusions}

\vskip 9pt

We have presented a simple model in 1+1 dimensions with a four fermion
 interaction term. Classically, the interaction term seemed to
violate Lorentz invariance. We have shown that, quantum mechanically,
 for  values of the
coupling constants satisfying the master equation, the model is not
only Lorentz invariant, but it is conformally
invariant as well.
 Apart from the intrinsic interest of these models,
 this opens the possibility
of utilizing them for string compactification.

\newpage

{\bf References}
\begin{enumerate}
\item M.B.Green, J.H.Schwarz and E.Witten, \emph{Superstring Theory},
Cambridge University Press, 1987.
\item J.Polchinski, \emph{String Theory}, Cambridge University Press,
 1998.
\item K.Bardakci and M.B.Halpern, Phys.Rev. {\bf D3}(1971) 2493.
\item M.B.Halpern and E.Kiritsis, Mod.Phys.Lett. {\bf A4}(1989) 1373;
Erratum ibid. {\bf A4} 1797. For a review of more recent developments,
see M.B.Halpern, E.Kiritsis, N.A.Obers and K.Clubok, Phys.Rept. {\bf 265}
(1996)1.
\item C.R.Nappi and E.Witten, Phys.Rev.Lett. {\bf 71}(1993) 3751.
\item M.B.Halpern and J.P.Yamron, Nucl.Phys. {\bf B351}(1991) 333.
\item J.de Boer, K.Clubok and M.B.Halpern, Int.J.Mod.Phys. {\bf A9}
(1994) 2451
\item A.A.Tseytlin, Nucl.Phys. {\bf B411}(1994) 509.
\item J.Bagger, D.Nemeschansky, N.Seiberg and S.Yankielovich,
Nucl.Phys. {\bf B289}(1987) 53.
\item R.Dashen and Y.Frishman, Phys.Rev. {\bf D11}(1975) 2781.
\item K.Bardakci and L.M.Bernardo, Nucl.Phys. {\bf B450}(1995) 695.
\item A.A.Tseytlin,Nucl.Phys. {\bf B418}(1994) 173.
\item O.O.Soloviev, Mod.Phys.Lett. {\bf A9}(1994) 483. 
\end{enumerate}
\end{document}